\documentclass[prd,preprint,superscriptaddress,tightenlines,nofootinbib, eqsecnum]{revtex4-2}

\usepackage{amsmath}
\usepackage{amsfonts}
\usepackage{amssymb}
\usepackage{bm}
\usepackage[colorlinks]{hyperref}
\usepackage{mathrsfs}
\usepackage{graphicx}
\usepackage{empheq}
\usepackage{ulem}
\usepackage{tensor}
\usepackage[usenames]{color}
\definecolor{darkgreen}{rgb}{0,0.5,0}
\hypersetup{urlcolor=darkgreen}
\usepackage[capitalize]{cleveref}

\allowdisplaybreaks

\DeclareSymbolFontAlphabet{\mathrsfs}{rsfs}
\DeclareMathAlphabet{\mathcal}{OMS}{cmsy}{m}{n}

\newcommand{\calO}{\mathcal{O}}
\newcommand{\dd}{\mathrm{d}}
\newcommand{\di}{\mathrm{i}}

\newcommand{\Mpl}{M_\mathrm{Pl}}

\newcommand{\tmT}{\tilde{\mu}_\text{T}^2}
\newcommand{\tG}{\tilde{\Gamma}_0}

\begin{document}
	
\title{Gravitational ringdown in the Minimal Theory of Massive Gravity}

\author{Hugo \textsc{Roussille}}\email{hugo.roussille@ens-lyon.fr}
\affiliation{Univ Lyon, ENS de Lyon, CNRS, Laboratoire de Physique, F-69342 Lyon, France}

\author{Fran\c{c}ois \textsc{Larrouturou}}\email{francois.larrouturou@obspm.fr}
\affiliation{
SYRTE, Observatoire de Paris, Université PSL, CNRS, Sorbonne Université, LNE, \\ 
61 avenue de l’Observatoire, 75014 Paris, France}

\date{\today}

\begin{abstract}
This work focuses on gravitational perturbations of black holes in the self-accelerating branch of the Minimal Theory of Massive Gravity (MTMG). This theory is a healthy extension of GR which displays the feature of massive tensor modes, without additional polarizations, strong-coupling issues nor requiring screening mechanisms. We proceed by implementing a newly developed technique that, instead of considering a second-order Schrödinger-like reformulation of perturbation equations, relies on a first-order formulation and solves it asymptotically, before numerically deriving the quasi-normal modes.
We find that the black holes of MTMG are stable, and that their quasi-normal modes smoothly differ from the GR ones, for non-vanishing values of the graviton mass.
This work hence confirms the fact that GR is a smooth limit of MTMG, and opens the exciting possibility of a clean test, performed for instance by the LISA detector.
\end{abstract}

\maketitle

\section{Introduction}
\label{sec_introduction}

In an era of precision gravitational astronomy, accurate waveform modeling is a milestone of signal analysis.
To the notable exception of the observation of binaries of neutron stars~\cite{LIGOScientific:2017vwq,LIGOScientific:2020aai}, the currently operating gravitational wave (GW) detectors are mainly sensitive to the instant of coalescence of binary systems~\cite{LIGOScientific:2021djp}.
Nevertheless, future generations of instruments, such as the Laser Interferometer Space Antenna (LISA)~\cite{LISA:2017pwj} or the Einstein Telescope (ET)~\cite{Branchesi:2023mws} are expected to be sensitive to the whole waveform.
For massive objects, the LISA instrument will notably be able to make clear observation of the \emph{ringdown} phase~\cite{Berti:2005ys,Berti:2016lat}, \emph{i.e.} the post-merger relaxation towards a single, stable compact object.
Such phase is driven by the so-called quasi-normal modes (QNM), typical damping frequencies from which one can extract valuable physical information on the remnant, as well as on the underlying gravitational theory.
In General Relativity (GR), for instance, where the \emph{isospectrality} theorem~\cite{Chandrasekhar:1985kt} imposes that, in vacuum and for asymptotically flat spacetimes, the QNMs are the same for both polarizations, many overtones have been computed, to impressively high accuracy, see \emph{e.g.}~\cite{Nollert:1999ji,Konoplya:2004uk,Berti:2009kk}.
In order to prepare the LISA mission, is it thus of prime importance to: (i) sharpen our predictions within GR; (ii) take into account environmental effects (such a accretion disks or third body interactions), that could induce departures from GR predictions; (iii) investigate the impact of possible deviations from GR.
This works follows the third direction, and aims at deriving quasi-normal modes within a stable and cosmologically relevant massive gravity framework.

Giving a mass to the graviton is one of the simplest conceptual extensions of GR, and has been implemented at the linear level as soon as 1939~\cite{Fierz:1939zz,Fierz:1939ix}.
Nevertheless, its non-linear completion reveals various subtleties~\cite{vanDam:1970vg,Zakharov:1970cc,Boulware:1972yco,Vainshtein:1972sx,Babichev:2009us,Babichev:2013usa}, and was only achieved seventy years later, with the de Rham-Gabadadze-Tolley (dRGT) model~\cite{deRham:2010ik,deRham:2010kj}, see~\cite{deRham:2014zqa} for a review on massive gravity.
Sadly, the dRGT theory suffers from phenomenological pathologies as it lacks stable homogeneous and isotropic cosmologies~\cite{DeFelice:2012mx} and the Ricci-flat black hole solutions of GR are linearly unstable~\cite{Babichev:2013una,Babichev:2014oua,Brito:2013wya}, which can seem problematic for the generation of gravitational radiation.
In addition, it propagates one scalar and two vector degrees of freedom on top of the two canonical tensor ones.
Those additional polarizations have not been detected to date~\cite{LIGOScientific:2019fpa,Takeda:2020tjj} (note however that those tests are to be taken with care, as they only contrast pure tensor \emph{vs.} pure scalar or pure vector scenarii, whereas dRGT predicts a mixing of all polarizations).
In order to solve those puzzles, frameworks completing dRGT gravity have been developed, see \emph{e.g.}~\cite{Hassan:2011zd,DAmico:2012hia,Huang:2012pe,DeFelice:2017gzc}.
Among those, a ``Minimal Theory of Massive Gravity'' (MTMG) has been proposed in~\cite{DeFelice:2015hla,DeFelice:2015moy} (and extended in~\cite{DeFelice:2022mcd}), namely a Lorentz symmetry breaking theory of massive gravity, propagating only two tensor polarizations (hence the name), which cosmology is free from instabilities and has rich phenomenological implications~\cite{Bolis:2018vzs,Hagala:2020eax,deAraujo:2021cnd,DeFelice:2021trp,DeFelice:2023bwq}.
Note however that, just as in the case of dRGT, the cosmology of MTMG requires to introduce an \emph{ad hoc} fiducial scale factor.
In order to avoid this inelegant tool, MTMG has been extended to render the fiducial scale factor dynamical~\cite{DeFelice:2017wel,DeFelice:2017rli,DeFelice:2018vkc}, as well as completed into a ``Minimal Theory of Bigravity''~\cite{DeFelice:2020ecp}.

Nevertheless, the original MTMG theory is a valuable framework to study effects beyond GR.
Indeed, its \emph{self-accelerating} cosmological branch has been shown to exactly mimic the canonical $\Lambda$CDM framework, even at the level of cosmological perturbations, to the only difference of having massive tensor modes~\cite{DeFelice:2015moy}.
Vacuum and star solutions were constructed in this branch~\cite{DeFelice:2018vza}, and it turns out that the static black hole solutions are nothing but the usual Schwarzschild-de Sitter spacetimes, written in a generalized Painlev\'{e}-Gullstrand coordinate system. 
Therefore MTMG is a very convenient framework to study deviations from GR during the generation of gravitational radiation, as one can isolate the effect of a massive graviton, without being polluted by other polarizations, black holes' hairs or cosmological instabilities.
This is the main motivation of this work, which aims at studying the stability and QNMs of the black hole solutions that were constructed in~\cite{DeFelice:2018vza}.

To do so, we will not try to reproduce the \emph{tour de force} performed in the framework of the Minimal Theory of Bigravity~\cite{Minamitsuji:2023lvi}, where master equations were explicitly constructed for the perturbation of the Schwarzschild-de Sitter solution derived in~\cite{Minamitsuji:2022vfv}.
Instead, we follow the method developed in~\cite{Langlois:2021xzq,Roussille:2022vfa} and implemented notably in scalar-tensor theories~\cite{Langlois:2021aji,Langlois:2022eta,Langlois:2022ulw,Noui:2023ksf,Roussille:2023sdr}.
Rather than trying to express the equations of motion as a Schrödinger-like second-order differential equation, this method consists in reducing the problem to a set of coupled first-order differential equations, that are solved asymptotically at both horizon and spatial infinity.
Using those asymptotic behaviors as boundary conditions, one can then numerically solve for the QNMs.
Such method is quite handy in the case of theories with additional (but possibly non-propagating) degrees of freedom, as will be the case in this work.
Indeed, it does not require the usually cumbersome construction of master variables, as physical modes will be easily identified asymptotically, and clearly separated from non-propagating ones.

This work is organized as follows.
Sec.~\ref{sec_model} presents the framework of MTMG, as well as the black hole solution that will be perturbed.
The asymptotic behavior of the perturbations are derived in Sec.~\ref{sec_asymptotic}, and the numerical computation of QNMs, \emph{via} a continued-fraction method, is exposed in Sec.~\ref{sec_QNM}.
Finally, Sec.~\ref{sec_concl} summarizes the main methods and results of this work.
As computations in MTMG are quite heavy, we applied in App.~\ref{app_PG} our procedure to the usual black hole of GR, but written in Painlev\'{e}-Gullstrand coordinates, in order to provide the reader with a simple example that encapsulates the principle of the MTMG computation.
Last, but not least, App.~\ref{app_lengthy} presents expressions that were too lengthy to be displayed in the main text.

\section{Model description}
\label{sec_model}

\subsection{The Minimal Theory of Massive Gravity}
\label{subsec_MTMG}

This work investigates the stability and ringdown phase of black holes in the ``Minimal Theory of Massive Gravity'', that was constructed in~\cite{DeFelice:2015hla,DeFelice:2015moy} as the unique theory of massive gravity that propagates only two tensor degrees of freedom, is invariant under spatial rotations and has the same cosmological backgrounds as dRGT gravity (see~\cite{DeFelice:2022mcd} for the construction of theories relaxing the last assumption).
Hereafter is presented a very compact description of MTMG, highlighting mostly the points that will be of interest in this work.
We let the interested reader refer to~\cite{DeFelice:2015hla,DeFelice:2015moy} for a more detailed and comprehensive description of the construction and cosmology of MTMG.

\subsubsection{Lagrangian description of MTMG}

In a nutshell, MTMG has been constructed starting from dRGT gravity expressed in a specific gauge, and adding appropriate constraints at the Hamiltonian level to kill the unwanted degrees of freedom, thus explicitly breaking Lorentz invariance.
It thus contains\footnote{The conventions employed throughout this work are as follows: we work with a mostly plus signature; greek letters denote spacetime indices (running from 0 to 3) and latin ones, either purely spatial indices (running from 1 to 3) or spherical indices (running from 1 to 2), depending on the (clear enough) context; (anti-)symmetrizations are weighted, \emph{e.g.} $A_{(ij)} = (A_{ij} + A_{ji})/2$; $\mathbb{I}_n$ denotes the identity matrix of rank $n$.} a physical metric $g_{\mu\nu}$, a fiducial metric $f_{\mu\nu}$ (coming from the original dRGT theory), as well as four Lagrange multipliers $\{\lambda,\lambda^i\}$, introduced with the additional constraints.
In its metric formulation, it reads
\begin{equation}\label{eq_MTMG_action}
\mathcal{S}_\text{MTMG} = \mathcal{S}_\text{EH}\left[g\right] + \mathcal{S}_\text{mat}\left[\Psi_\text{m};g\right] - \frac{\Mpl^2\,m_g^2}{2}\int\!\!\dd^4x \sqrt{-g}\,\mathcal{W}\,,
\end{equation}
where $\mathcal{S}_\text{EH}$ is the usual Einstein-Hilbert action for the metric $g_{\mu\nu}$, $\mathcal{S}_\text{mat}$ is a minimally coupled matter action, $\Mpl$ is the Planck mass and $m_g$ is an additional mass scale.
In order to express the potential term $\mathcal{W}$, one has to perform an Arnowitt-Deser-Misner (ADM) foliation of the physical metric
\begin{equation}\label{eq_MTMG_ADMg}
\dd s^2 = g_{\mu\nu}\dd x^\mu \dd x^\nu 
= -N^2 \dd t^2 + \gamma_{ij}\big(N^i \dd t + \dd x^i\big)\big(N^j \dd t + \dd x^j\big)\,,
\end{equation}
as well as a similar decomposition of the fiducial one, defined with vanishing shift
\begin{equation}\label{eq_MTMG_ADMf}
f_{\mu\nu}\dd x^\mu \dd x^\nu 
= -\tilde{N}^2 \dd t^2 + \tilde{\gamma}_{ij}\,\dd x^i\,\dd x^j\,.
\end{equation}
From those foliations, one can define the square-root matrix $\mathfrak{K}^i_{\ j}$, as well as the extrinsic curvatures $K^i_{\ j}$ and $\tilde{\zeta}^i_{\ j}$ as
\begin{equation}
\mathfrak{K}^i_{\ k} \mathfrak{K}^k_{\ j} \equiv \gamma^{ik}\tilde{\gamma}_{kj}\,,
\qquad
K^i_{\ j}  \equiv \frac{1}{2N}\,\gamma^{ik}\bigg( \partial_t\gamma_{kj} - 2 \mathcal{D}_{(k}N_{j)}\bigg)\,,
\qquad
\tilde{\zeta}^i_{\ j} \equiv \frac{1}{2\tilde{N}}\,\tilde{\gamma}^{ik}\partial_t\tilde{\gamma}_{kj}\,,
\end{equation}
where $\gamma^{ij}$ and $\tilde{\gamma}^{ij}$ are inverse to the spatial metrics $\gamma_{ij}$ and $\tilde{\gamma}_{ij}$, and $\mathcal{D}_i$ is the covariant derivative compatible with $\gamma_{ij}$.
The square-root matrix $\mathfrak{K}^i_{\ j}$ enters the potential term $\mathcal{W}$ through the combinations
\begin{equation}
\begin{aligned}
&
\mathcal{E} \equiv \sum^3_{n=0} c_n e_{3-n}(\mathfrak{K})\,,
\qquad
& &
\tilde{\mathcal{E}} \equiv \sum^4_{n=1} c_n e_{4-n}(\mathfrak{K})\,,
\qquad
& &
\hat{\mathcal{E}} \equiv \sum^4_{n=2} c_n e_{5-n}(\mathfrak{K})\,,\\
& & &
\tilde{\mathcal{F}}^i_{\ j} \equiv \frac{\delta \tilde{\mathcal{E}}}{\delta \mathfrak{K}^j_{\ i}}\,,
& &
\hat{\mathcal{F}}^i_{\ j} \equiv \frac{\delta \hat{\mathcal{E}}}{\delta \mathfrak{K}^j_{\ i}}\,,
\end{aligned}
\end{equation}
where $e_n$ are the three-dimensional symmetric polynomials and $\{c_n\}_{n=0..4}$ are constants defining the theory (note that one of those constants can be absorbed in the scale $m_g$).
Finally, the potential term is written as
\begin{equation}\label{eq_MTMG_Wdef}
\mathcal{W} \equiv 
\frac{\tilde{N}}{N}\,\mathcal{E} + \tilde{\mathcal{E}}
+ \frac{\tilde{N}\,\lambda}{N}\bigg(\hat{\mathcal{F}}^i_{\ j}\tilde{\zeta}^j_{\ i} - \tilde{\mathcal{E}}\tilde{\zeta}^i_{\ i} + \tilde{\mathcal{F}}^i_{\ j}K^j_{\ i}\bigg)
+ \frac{\tilde{N}}{N}\,\tilde{\mathcal{F}}^i_{\ j}\mathcal{D}_i\lambda^j
- \frac{m_g^2\tilde{N}^2\lambda^2}{4N^2}\bigg( \tilde{\mathcal{F}}^i_{\ j} - \frac{\tilde{\mathcal{F}}^k_{\ k}}{2}\,\delta^i_j\bigg)\tilde{\mathcal{F}}^j_{\ i}\,,
\end{equation}
which defines uniquely MTMG.

\subsubsection{Cosmological phenomenology of MTMG}

Just as dRGT gravity does, MTMG enjoys two branches of homogeneous and isotropic cosmological background solutions.
In the so-called ``self-accelerating'' one, the ratio of the two scale factors (the fiducial $a_f$ and the physical $a$), $\mathcal{X} \equiv a_f/a$ is constant, the combination $c_1\mathcal{X}^2 + 2c_2\mathcal{X} + c_3$ vanishes, and the mass term~\eqref{eq_MTMG_Wdef} reduces to an effective cosmological constant.
Even at the level of the perturbations, both scalar and vector modes mimic exactly the phenomenology of the canonical $\Lambda$CDM framework.
As for the tensor sector, it is free from strong-coupling issues and propagates at the speed of light.
The main difference is that it bears a mass
\begin{equation}\label{eq_MTMG_muT2_cosmo}
\mu_\text{T,cosmo}^2 = \frac{m_g^2}{2}\,\mathcal{X}\big(c_1\mathcal{X}+c_2\big)\bigg(\frac{\tilde{N}}{N}-\mathcal{X}\bigg)\,,
\end{equation}
where we recall that $N$ and $\tilde{N}$ are the two lapses of the foliations~\eqref{eq_MTMG_ADMg}--\eqref{eq_MTMG_ADMf} and $c_{1,2}$ are constants defining the theory.
Note that, if $\mathcal{X}$ is constant, the ration $\tilde{N}/N$ is \emph{a priori} not, and thus the graviton's mass can be time-varying, which yields interesting phenomenological implications.
Naturally, one has to ensure the reality of the mass, $\mu_\text{T,cosmo}^2>0$, in order to avoid tachyonic instabilities (note also that, as the theory does not propagate gravitational scalar degrees of freedom, it is free from Higuchi-like ghosts).

In the second branch, dubbed ``normal'', $\mathcal{X}$ is no more constant and the mass term~\eqref{eq_MTMG_Wdef} gives rise to an effective dark fluid, whose equation of state evolves with time (but is only and uniquely determined by the matter content of the Universe).
This yields rich phenomenological implications, notably a modified growth of perturbations, see~\cite{Bolis:2018vzs,Hagala:2020eax,deAraujo:2021cnd,DeFelice:2021trp}.
Nevertheless this work focuses on the first branch, as it offers a quite convenient framework for testing minimal deviations from GR, as advertised in the introduction.

\subsection{Black hole solutions in MTMG}
\label{subsec_MTMG_BH}

As proven in~\cite{DeFelice:2018vza}, the MTMG enjoys a useful \emph{mimicking lemma}: any GR solution that can be written with flat constant-time surfaces is a solution of the self-
accelerating branch of MTMG, with the additional feature of a bare cosmological constant. 
This lemma allows to construct (static) black hole and star solutions in MTMG, by simply finding a coordinate system in which the usual GR solutions admit spatially flat foliations
Note an important consequence of the breaking of Lorentz symmetry in MTMG: for a given GR spacetime, two different foliations with flat constant-time surfaces correspond to two different \emph{physical} spacetimes.

It is long known that Schwarzschild spacetimes admit spatially flat coordinate system~\cite{Painleve,Gullstrand,Lemaitre}, and those can be further generalized to the case of Schwarzschild-de Sitter spacetimes of mass $M$ and cosmological constant $\Lambda$ as
\begin{equation}\label{eq_MTMG_ds2PG}
\dd s^2_\text{PG} = - \dd t^2 + \bigg( \dd r \pm \sqrt{\frac{2M}{r}- \frac{\Lambda r^2}{3}}\,\dd t\bigg)^2 + r^2 \dd \Omega^2\,,
\end{equation}
where $\dd \Omega^2$ is the usual 2-sphere volume element.
One advantage of those coordinates is that they are regular at the black-hole horizon, and cover the whole spacetime between the origin and the (possible) cosmological horizon.
Indeed, the presence of an event horizon does not come from a singular behavior of the metric but from the fact that the light-cone's tangent intersects the $r = $ constant surface for $\sqrt{\frac{2M}{r}- \frac{\Lambda r^2}{3}} = 1$.
It is straightforward to see that the $+$ sign of the line element~\eqref{eq_MTMG_ds2PG} corresponds to a black hole, whereas the $-$ one corresponds to a white hole.

By virtue of the mimicking lemma, and as explicitely shown in~\cite{DeFelice:2018vza}, the line-element~\eqref{eq_MTMG_ds2PG} is a solution of the self-accelerating branch of MTMG in vacuum, with the identification
\begin{equation}
\Lambda = \Lambda_\text{eff} = \frac{m_g^2}{2}\bigg(c_4 + 2c_ 3\mathcal{X} + c_2 \mathcal{X}^2\bigg)\,,
\end{equation}
where $\mathcal{X}$ is a constant satisfying $c_1\mathcal{X}^2 + 2c_2\mathcal{X} + c_3=0$, and the fiducial and auxiliary sectors are given by
\begin{equation}\label{eq_MTMG_BH_fid_aux}
\tilde{N} = 1 \,,
\qquad
\tilde{\gamma}_{ij}\dd x^i \dd x^j = \mathcal{X}^2 \bigg[ \dd r^2 + r^2 \dd \Omega^2\bigg]\,,
\qquad
\lambda = \lambda^i = 0\,.
\end{equation}
This black hole solution is the one that will be perturbed hereafter.

\section{Asymptotic behavior of the perturbations}
\label{sec_asymptotic}

Instead of seeking for Schrödinger-like, second-order differential master equations, we follow the formalism developed in~\cite{Langlois:2021xzq,Roussille:2022vfa}, relying on first-order systems.\footnote{Analytic and numeric computations presented in this work were performed by relying on the \emph{Mathematica} software~\cite{Mathematica}.}
This procedure consists in two steps: (i) investigating the behavior of the perturbations at both spatial infinity and horizon by solving the equations of motion in an asymptotic perturbative expansion; (ii) using those boundary behaviors to numerically derive the QNMs.
This section deals with the first step, and the next Sec.~\ref{sec_QNM}, with the second one.
For the sake of simplicity, we restrain this work to the case of an asymptotically flat background spacetime, namely the line element~\eqref{eq_MTMG_ds2PG} with $\Lambda = 0$ together with the fiducial and auxiliary sectors~\eqref{eq_MTMG_BH_fid_aux}.

We let the reader refer to App.~\ref{app_PG} for a detailed application of the formalism on the spacetime~\eqref{eq_MTMG_ds2PG}, but within the simpler framework of GR.

\subsection{Reduction to a first-order system}
\label{subsec_asymptotic_reduction}

The first step is naturally to perturb the black hole solution~\eqref{eq_MTMG_ds2PG} and the Lagrange multipliers $\{\lambda,\lambda^i\}$.
The fiducial metric being fiducial, it will not be perturbed.
As the background spacetime is static and spherically symmetric, the perturbations can be expanded in spherical harmonics, and decomposed on a time-Fourier basis.
We introduce $\omega_{ab}$ as the usual 2-sphere metric, $\dd\Omega^2 = \omega_{ab}\,\dd \zeta^a \dd \zeta^b = \dd \theta^2 + \sin^2\theta \,\dd \varphi^2$, with compatible covariant derivative $\nabla_a$, $Y_L^M(\zeta^a)$ a basis of spherical harmonics and the weighted anti-symmetric symbol $\tilde{\varepsilon}_{ab} = \sqrt\omega \,\varepsilon_{ab}$.
In the spirit of~\cite{Regge:1957td}, we define the metric perturbations by their behavior under the 2-sphere rotation symmetry group, as
\begin{subequations}\label{eq_asym_pertmetric}
\begin{align}
&
N = 1 + e^{-\di \omega t}\,n^0_{\ell,m}(r)\,Y_L^M\,,\\
&
N^r =\sqrt{\frac{2M}{r}} + e^{-\di \omega t}\,n^r_{\ell,m}(r)\,Y_L^M\,,\\
&
N^a = e^{-\di \omega t}\,\bigg[n^+_{\ell,m}(r)\,\partial^a +  n^-_{\ell,m}(r)\,\tilde{\varepsilon}^{ab}\,\partial_b\bigg]Y_L^M \,,\\
&
\gamma_{rr} = 1 + e^{-\di \omega t}\,h^0_{\ell,m}(r)\,Y_L^M \,,\\
&
\gamma_{ra} = e^{-\di \omega t}\,\bigg[h^+_{\ell,m}(r)\,\partial_a +  h^-_{\ell,m}(r)\,\tilde{\varepsilon}_{ab}\,\partial^b\bigg]Y_L^M \,,\\
&
\gamma_{ab} = r^2\omega_{ab}+e^{-\di \omega t}\,\bigg[k^0_{\ell,m}(r)\,\omega_{ab} + k^+_{\ell,m}(r)\,\nabla_{ab} - k^-_{\ell,m}(r)\,\tilde{\varepsilon}_{c(a}\nabla^c\partial_{b)}\bigg] Y_L^M \,.
\end{align}
\end{subequations}
Similarly, we perturb the Lagrange multipliers as
\begin{subequations}\label{eq_asym_pertaux}
\begin{align}
&
\lambda = e^{-\di \omega t}\,\ell^0_{\ell,m}(r)\,Y_L^M\,,\\
&
\lambda^r = e^{-\di \omega t}\,\ell^r_{\ell,m}(r)\,Y_L^M\,,\\
&
\lambda^a = e^{-\di \omega t}\,\bigg[\ell^+_{\ell,m}(r)\,\partial^a +  \ell^-_{\ell,m}(r)\,\tilde{\varepsilon}^{ab}\,\partial_b\bigg]Y_L^M \,.
\end{align}
\end{subequations}
For the sake of lightness, we will stop writing the $(\ell,m)$ dependence of the modes, as different $(\ell,m)$ modes do not interact at the linear level.
It is easy to see that our theory contains ten even parity modes (denoted with 0, $r$ or $+$ exponents) and four odd parity modes (denoted with $-$ exponents).
As MTMG is invariant under the parity transformation, the even and odd modes decouple at linear order, and we can treat each sector independently.
Injecting the parametrization~\eqref{eq_asym_pertmetric}--\eqref{eq_asym_pertaux} into the action~\eqref{eq_MTMG_action} in vacuum (\emph{i.e.} taking $\mathcal{S}_\text{mat} = 0$) and varying it, we end as expected with ten even parity and four odd parity equations of motion.
Note here that, due to the explicitly breaking the diffeomorphism invariance, we cannot implement the usual gauge choices~\cite{Regge:1957td,Zerilli:1970se} reducing the number of variables, and thus we need to consider the full set of equations.\footnote{If we had perturbed the fiducial spatial metric too, we could have chosen a gauge within the class of \emph{foliation-preserving} diffeomorphism invariance (see \emph{e.g.} Eq.~(12) of~\cite{DeFelice:2020onz}), thus effectively killing three degrees of freedom, for instance $\{h^+,k^+,k^-\}$. Nevertheless this procedure amounts to introducing Stueckelberg fields at the level of the theory, which we wanted to avoid.}
In the following, we focus on propagating modes\footnote{A rapid inspection of the non-propagating modes $\ell = 0,1$ revealed that they are healthy, hence the following treatment concentrates on propagating ones.} $\ell \geq 2$ and we use adimensioned variables, $\rho = r/(2M)$ and $\tilde{\omega} = 2M\omega$ together with the combinations
\begin{equation}\label{eq_asym_muT_Gamma0_def}
\tmT = \frac{m_g^2}{8M^2}\,\mathcal{X}\big(1-\mathcal{X}\big)\big(c_1\mathcal{X}+c_2\big)\,,
\qquad
\tG = \frac{m_g^2}{4M^2}\,\mathcal{X}\big(c_0\mathcal{X}^2+2c_1\mathcal{X}+c_2\big)\,.
\end{equation}
Note that, up to the adimensionalization $1/(2M)^2$, $\tmT$ coincides with the cosmological mass of tensor modes~\eqref{eq_MTMG_muT2_cosmo}.

\subsubsection{Odd parity sector}

Among the four odd parity equations of motion, two of them are second-order differential equations.
Using the usual trick to reduce them\footnote{Namely, for a second-order differential equation $y'' + a y' + by = c$, we introduce $\tilde{y} = y'$ so that it is equivalent to the coupled first-order system $\{y' - \tilde{y} = 0, \tilde{y}' + a \tilde{y} + b y = c\}$.} we reach a system of six coupled first-order differential equations of six variables, which derivative sector can be further diagonalized to yield the sought form 
\begin{equation}\label{eq_asym_1storder_odd}
\frac{\dd X_\text{odd}}{\dd \rho} + B_\text{odd} \cdot X_\text{odd} = 0\,,
\end{equation}
where the relation between $X_\text{odd}$ and the original metric perturbations, as well as the explicit expression of $B_\text{odd}$ are given in App.~\ref{app_lengthy_1storder_odd}.
Note that among the MTMG-specific combinations~\eqref{eq_asym_muT_Gamma0_def}, only $\tmT$ appears.

\subsubsection{Even parity sector}

Among the ten odd parity equations of motion, four of them are second-order differential equations, which are reduced to yield a system of fourteen coupled equations of (at most) first order.
Fortunately, four combinations of those fourteen equations are independent algebraic relations, that can be implemented to remove four variables.
Thus, we find a system of ten coupled first-order differential equations of ten variables, whose derivative sector can be further diagonalized to yield
\begin{equation}\label{eq_asym_1storder_even}
\frac{\dd X_\text{even}}{\dd \rho} + B_\text{even} \cdot X_\text{even} = 0\,,
\end{equation}
where the relation between $X_\text{even}$ and the original metric perturbations, as well as the explicit expression of $B_\text{even}$ are given in App.~\ref{app_lengthy_1storder_even}.
Note that both $\tmT$ and $\tG$ appear in those expressions.

\subsection{Odd parity modes}
\label{subsec_asymptotic_odd}

Let us first focus on the simpler case, the odd parity modes.
The aim is to perturbatively solve the first-order system~\eqref{eq_asym_1storder_odd} at both spatial infinity $\rho \to \infty$ and horizon $\rho\to1$.
This would be trivially achieved if the matrix $B_\text{odd}$ were to be diagonal, so our aim is to asymptotically diagonalize it.
Not all six variables are expected to propagate, thus the diagonal matrix will contain components corresponding to both propagating and non-propagating modes, and we will need to distinguish them.
Note that, due to the possible $\rho$ dependency of the transition matrix $P$, the new variable $Y = P^{-1}X_\text{odd}$ obeys
\begin{equation}\label{eq_asym_odd_transfo_rule}
\frac{\dd Y}{\dd \rho} + \tilde{B}\cdot Y = 0\,,
\qquad\text{where}\qquad
\tilde{B} = P^{-1}\cdot B_\text{odd}\cdot P + P^{-1}\cdot\frac{\dd P}{\dd \rho}\,.
\end{equation}

\subsubsection{Behavior at spatial infinity}
\label{subsubsec_asymptotic_odd_infty}

Let us first start with the study a spatial infinity and use the variable $\varrho = \sqrt{\rho}$ for simplicity.
Taylor expanding the system~\eqref{eq_asym_1storder_odd} around $\varrho \to \infty$ by using the explicit expression for the matrix $B_\text{odd}$~\eqref{eq_app_Bodd}, it comes
\begin{equation}\label{eq_asym_odd_infty_expanded}
\frac{\dd X_\text{odd}}{\dd \varrho} + \left(\varrho\,B^{\infty,(-1)}_\text{odd}+ \sum_{n=0}^\infty\,\frac{1}{\varrho^n}\,B^{\infty,(n)}_\text{odd}\right)\cdot X_\text{odd} = 0\,.
\end{equation}
The dominant order, $B^{\infty,(-1)}_\text{odd}$ has two non-vanishing eigenvalues, $\mp 2\di \sqrt{\tilde{\omega}^2-\tmT}$, corresponding to the expected dispersion relation of massive waves at infinity (the factor 2 comes from the square-root in the definition of $\varrho$).
The aim is now to diagonalize the system~\eqref{eq_asym_odd_infty_expanded} to a given precision, say $N$.
To do so, we proceed in two steps.
First, we diagonalize it by block, seeking to decouple the previsouly found eigenvalues $\mp 2\di \sqrt{\tilde{\omega}^2-\tmT}$ from the rest of the system.
This is achieved by using an ansatz of the form
\begin{equation}
P_\infty = P_\infty^0 \cdot \left[\mathbb{I}_6 + \sum_{n=1}^N\frac{1}{\varrho^n}\, P_\infty^{(n)}\right]\,,
\quad\text{where}\quad
P_\infty^{(n)} =
\left(
\begin{array}{c|c}
\begin{matrix}
0 & \alpha_{1,1}^{(n)} \\
\beta_{1,1}^{(n)} & 0
\end{matrix}
&
\alpha_{4,2}^{(n)}\\
\hline
\beta_{2,4}^{(n)} & 0_{4,4}
 \end{array}\right)
\end{equation}
and $P_\infty^0$ is the transition matrix associated with the Jordan decomposition of $B^{\infty,(-1)}_\text{odd}$.
As explained in details in~\cite{Langlois:2021xzq,Roussille:2022vfa}, this ansatz transforms a complex diagonalization process into a simpler linear problem.
Indeed, using the transformation rule~\eqref{eq_asym_odd_transfo_rule} one can see that at a given order, the matrix $P_\infty^{(n)}$ should simply obey $[A_1,P_\infty^{(n)}] = A_2$, where $A_1$ and $A_2$ are matrices composed of products of already known $B^{\infty,(n)}_\text{odd}$ and $P_\infty^{(m<n)}$.
Such a problem can thus be solved to any precision, and we obtain a system of the type
\begin{equation}
\frac{\dd Y}{\dd \varrho} = \Bigg[\varrho\, 
\begin{pmatrix}
\Pi_2^\infty & 0\\
0 & \Sigma_4^\infty
\end{pmatrix}+ \mathcal{O}\left(\varrho^{-N}\right)
\Bigg]\cdot Y \,,
\end{equation}
where $\Pi_2^\infty$ is a diagonal matrix with leading order given by $\pm 2\di \sqrt{\tilde{\omega}^2-\tmT}$, and $\Sigma_4^\infty$ is a complicated $4\times 4$ matrix.
The next step is thus to diagonalize $\Sigma_4^\infty$, whose leading order is nilpotent.
This nilpotence simply translates a non-optimal choice of variables, and can be removed by applying the simple transition matrix $\text{diag}[1,1,1,\varrho^2,1,\varrho^{-2}]$.
After this operation, it appears that this sector starts at order $\varrho^{-2}$ and can be easily diagonalized order by order.
In those new variables, the system becomes
\begin{equation}\label{eq_asym_odd_infty_diag}
\frac{\dd Z}{\dd \varrho} = \bigg[ \varrho\, 
D_\text{odd}^\infty  + \mathcal{O}\left(\varrho^{-N}\right)
\bigg]\cdot Z\,,
\end{equation}
where $D_\text{odd}^\infty$ is now a diagonal matrix, whose components $\{\pi^\infty_\pm,\sigma_\pm,\varsigma_\pm\}$ are explicitly given by
\begin{subequations}
\begin{align}
&\label{eq_asym_odd_infty_pi}
\pi^\infty_\pm = \pm 2\di\,\sqrt{\tilde{\omega}^2-\tmT} + \frac{2\di\,\tilde{\omega}}{\varrho} + \frac{1}{\varrho^2}\bigg[2\pm\di\, \frac{2\tilde{\omega}^2-\tmT}{\sqrt{\tilde{\omega}^2-\tmT}}\bigg] + \mathcal{O}\left(\frac{1}{\varrho^3}\right)\,,\\
&
\sigma_\pm = \pm\frac{2\ell}{\varrho^2} + \mathcal{O}\left(\frac{1}{\varrho^5}\right)\,,
\qquad\qquad
\varsigma_\pm = \pm\frac{2(\ell+1)}{\varrho^2} + \mathcal{O}\left(\frac{1}{\varrho^5}\right)\,.
\end{align}
\end{subequations}
Integrating the system~\eqref{eq_asym_odd_infty_diag} and going back to the primary $\rho$ coordinate, it is easy to isolate the two propagating modes, corresponding the the values $\pi^\infty_\pm$,
\begin{equation}\label{eq_asym_odd_z_infty}
z_\pm^\text{prop} \sim e^{\pm \di\,\sqrt{\tilde{\omega}^2-\tmT}\,\rho + 2\di\,\tilde{\omega}\sqrt{\rho}} \,\rho^{1\pm\di\, \frac{2\tilde{\omega}^2-\tmT}{2\sqrt{\tilde{\omega}^2-\tmT}}}\,,
\end{equation}
from the four non-propagating modes
\begin{equation}
z^\text{static}_n \sim \rho^{\lambda_n}\,,
\qquad\text{with}\qquad \lambda_n = \big\lbrace\pm\ell,\pm(\ell+1)\big\rbrace\,,
\end{equation}
which bear the radial dependencies of usual static multipolar modes.
Note that, in the $\tmT \to 0$ limit, the propagating modes~\eqref{eq_asym_odd_z_infty} behave exactly like the GR modes~\eqref{eq_app_Xinf}, which is a promising sign of stability.
Notably, the unusual dependence $e^{2\di\,\tilde{\omega}\sqrt{\rho}}$ is simply due to the square-root appearing in the PG coordinate system.

One can then invert the transition matrices to express $X_\text{odd}$ in terms of $Z$, and use the relation~\eqref{eq_app_Xodd_func_h} to express the asymptotic behavior of the metric perturbations, displayed in Eq.~\eqref{eq_app_asym_odd_infty}.
It is important to remark that the auxiliary field $\ell^-$ does not propagate at infinity (this fact has been tested up to the precision $N = 500$).

\subsubsection{Behavior at horizon}\label{subsubsec_asymptotic_odd_hor}

At horizon, let us expand the system~\eqref{eq_asym_1storder_odd} in terms of the variable $\theta = (\rho-1)^{-1}$
\begin{equation}\label{eq_asym_odd_infty_expanded}
\frac{\dd X_\text{odd}}{\dd \theta} + \left( \sum_{n=1}^\infty\,\frac{1}{\theta^n}\,B^{\text{H},(n)}_\text{odd}\right)\cdot X_\text{odd} = 0\,.
\end{equation}
Interestingly, the dominant order, $B^{\text{H},(1)}_\text{odd}$ has only one non-vanishing eigenvalue, $2\di\,\tilde{\omega}-1$.
Using the same technique as the one used for the study at infinity, one can diagonalize the system by block, as
\begin{equation}
\frac{\dd \hat{Y}}{\dd \theta} = \Bigg[ \frac{1}{\theta}\, 
\begin{pmatrix}
\pi_+^\text{H} & 0\\
0 & \frac{1}{\theta} \Sigma_5^\text{H}
\end{pmatrix}
+\mathcal{O}\left(\theta^{-N-2}\right)
\Bigg]\cdot \hat{Y}\,,
\end{equation}
where
\begin{equation}\label{eq_asym_odd_infty_piH}
\pi_+^\text{H} = 1-2\di\,\tilde{\omega} + \frac{2(\ell^2+\ell-3)+5\di\,\tilde{\omega}-2(3\tilde{\omega}^2-\tmT)}{2(1-2\di\,\tilde{\omega})\theta}+\mathcal{O}\left(\frac{1}{\theta^2}\right)\,,
\end{equation}
and $\Sigma_5^\text{H}$ is a complicated $5\times 5$ matrix.
At leading order, its determinant
\begin{equation}
\text{det}\Sigma_5^\text{H} = \frac{1}{2\di\,\tilde{\omega}-1}\bigg\lbrace
\ell(\ell+1)\bigg[\big(\ell^2+\ell+2\big)\big(\tilde{\omega}^2-\tmT\big)+10\di\,\tilde{\omega} - \ell(\ell+1)\big(\ell^2+\ell-1\big)\bigg]-12\bigg\rbrace
\end{equation}
is clearly non-vanishing, and thus $\Sigma_5^\text{H}$ is invertible.
For generic $\{\ell,\tilde{\omega},\tmT\}$, we were not able to write the associated transition matrix in closed form, but rather as sum of roots of fifth order polynomials.
Note however that, once values for $\{\ell,\tilde{\omega},\tmT\}$ have been chosen, the transition matrix is easily derived.
Nevertheless this transition matrix exists and so there is a transformation such that the system becomes
\begin{equation}\label{eq_asym_odd_hor_diag}
\frac{\dd \hat{Z}}{\dd \theta} = \bigg[ \frac{1}{\theta}\, 
D_\text{odd}^\text{H} + \mathcal{O}\left(\theta^{-N-2}\right) \bigg]\cdot \hat{Z}\,,
\end{equation}
where $D_\text{odd}^\text{H}$ is now a diagonal matrix, with components $\{\pi^\text{H}_+,\sigma_n/\theta\}_{n=2..6} + \mathcal{O}(\theta^{-2})$.
Integrating it and returning to the original radial coordinate, one has
\begin{subequations}
\begin{align}
&\label{eq_asym_odd_hor_zoutgoing}
z_1 \sim \big(\rho-1\big)^{2\di\,\tilde{\omega}-1}\,e^{\frac{2(\ell^2+\ell-3-3\tilde{\omega}^2+\tmT)+5\di\,\tilde{\omega}}{2(1-2\di\,\tilde{\omega})}(\rho-1)}\,,\\
&\label{eq_asym_odd_hor_zn}
z_n \sim e^{\sigma_n(\rho-1)}\,,\hspace{6cm}\text{for}\quad n=2..6\,.
\end{align}
\end{subequations}
Using Eddington–Finkelstein-like coordinate $\rho_\star =\ln(\rho-1)$, one can identify $z_1$ with the outgoing mode, although its dispersion relation $\vert \vec{k}\vert=2\omega$ is not canonical.
What about the ingoing mode ?
Going to large $\ell$, one can diagonalize the leading order of $\Sigma_5^\text{H}$ and we find the following five eigenvalues
\begin{subequations}
\begin{align}
&\label{eq_asym_odd_hor_z1}
\sigma_2 =\di\,\tilde{\omega} -\frac{1}{2} - \frac{2(\ell^2+\ell-3)+5\di\,\tilde{\omega}-2(3\tilde{\omega}^2-\tmT)}{2(1-2\di\,\tilde{\omega})}+ \mathcal{O}\left(\frac{1}{\lambda}\right)\,,\\
&
\sigma_{3,4} = \pm \sqrt{2(2+\lambda)} + \mathcal{O}\left(\frac{1}{\lambda}\right)\,,\\
&
\sigma_{5,6} = \pm \sqrt{2\lambda} + \frac{5}{4} \pm \frac{49- 40\di\,\tilde{\omega}}{32\,\sqrt{2\lambda}}+ \mathcal{O}\left(\frac{1}{\lambda}\right)\,,
\end{align}
\end{subequations}
where $\lambda = \ell(\ell+1)/2-1$.
Only the eigenvalue $\sigma_2$ corresponds to a (ingoing) propagating mode (note that the last term matches the subleading behavior of $z_1$~\eqref{eq_asym_odd_hor_z1}, up to a global sign).
Although not trivial to see \emph{a priori}, the system therefore contains the expected modes.

This non-canonical behavior at horizon (constant ingoing mode, non-usual dispersion relation for the outgoing one) is nonetheless \emph{not} a feature of the theory under investigation, but rather of the coordinate system.
Indeed, the study of a black hole in GR, written in the same PG coordinates, reveals the same phenomenology, as shown in Eq.~\eqref{eq_app_Xhor}.
The main difference between the usual Schwarzschild coordinates and the PG ones is the fact that the latter are perfectly regular at horizon.
The ``singular'' nature of the horizon is encrypted in the behavior of the light-cone structure: at horizon, its radial outgoing tangent is directed along the $r = $ constant surface, which make the horizon a trapped surface.
Therefore outgoing waves bear a mild singularity~\eqref{eq_asym_odd_hor_zoutgoing}, and enjoy a non-canonical relation dispersion.
As for ingoing waves, the corresponding light-cone tangent simply obeys $(\dd t/\dd r)_{r=2M} = -1/2$ and there is no coordinate singularity: the horizon is not a particularly relevant point of spacetime and hence the modes behave smoothly as~\eqref{eq_asym_odd_hor_zn}, with $\sigma_2$ being given by~Eq.~\eqref{eq_asym_odd_hor_z1}.


\subsection{Even parity modes}
\label{subsec_asymptotic_even}

The treatment of the even parity sector~\eqref{eq_asym_1storder_even} is conceptually identical to the treatment of the odd parity one, the main technical difference being the larger size of the system, increasing the complexity of the computations.

As discussed herebelow, it only amounts to adding static modes: the propagating modes match exactly their odd sector counterparts, and have the same dispersion relations.
Although definitively not a formal proof, this could point out that an isospectrality theorem may exist in MTMG.
Nevertheless, we have not tried to derive such a proof, neither have we computed the QNMs of the even sector (which may have brought a counter-example), so we let the investigation of isospectrality in MTMG for future work.

\subsubsection{Behavior at spatial infinity}

Following the lines of Sec.~\ref{subsubsec_asymptotic_odd_infty}, we use the variable $\varrho = \sqrt{\rho}$ to Taylor expand the system~\eqref{eq_asym_1storder_even} at spatial infinity, and diagonalize it asymptotically up to order $N$, as
\begin{equation}\label{eq_asym_even_infty_diag}
\frac{\dd Z_\text{even}}{\dd \varrho} = \bigg[ \varrho\,D^\infty_\text{even} + \mathcal{O}\left(\varrho^{-N}\right)
\bigg]\cdot Z_\text{even}\,,
\end{equation}
where $D^\infty_\text{even}$ is a diagonal matrix of order 10.
Among its components, two correspond to propagating modes and eight, to static ones.
The propagating ones exactly match the propagating modes from the odd sector $\pi_\pm^\infty$~\eqref{eq_asym_odd_infty_pi}.
As for the static modes, four of them are independent of the parameter $\tG$, and the remaining four have a complex structure involving $\tG$,
\begin{subequations}
\begin{align}
&
\hat{\sigma}_{1..4} = \big(\ell+2,\ell+1,1-\ell,-\ell\big)\,,\\
&
\hat{\sigma}_{5..8} = -1 \pm \sqrt{\frac{1}{2}+\ell(\ell+1) \pm \sqrt{\tilde{\gamma}_0}}\,,
\end{align}
\end{subequations}
where we have shortened
\begin{equation}
\tilde{\gamma}_0 = \frac{1}{4}+\ell(\ell+1) - \frac{9(\ell-1)\ell(\ell+1)(\ell+2)}{2(\tilde{\omega}^2-\tmT)}\,\tG\,.
\end{equation}
However, the components $\hat{\sigma}_{5..8}$ nicely reduce to $\hat{\sigma}_{1..4}$ when setting $\tG = 0$ (as $\ell\geq 2$).

Inverting the transition matrices and returning to the original metric perturbations, one can see that, contrarily to the odd sector, all perturbations propagate.

\subsubsection{Behavior at horizon}

Expanding the system~\eqref{eq_asym_1storder_even} in terms of the variable $\theta = (\rho-1)^{-1}$ and diagonalizing it by block, one reaches
\begin{equation}\label{eq_asym_even_hor_diag}
\frac{\dd \hat{Y}_\text{even}}{\dd \theta} = \Bigg[ \frac{1}{\theta}\, 
\begin{pmatrix}
\pi_+^\text{H} & 0\\
0 & \frac{1}{\theta} \Sigma_9^\text{H}
\end{pmatrix}
+\mathcal{O}\left(\theta^{-N-2}\right)
\Bigg]\cdot \hat{Y}_\text{even}\,,
\end{equation}
where $\pi_+^\text{H}$ is the same as in the odd sector, Eq.~\eqref{eq_asym_odd_infty_piH}, and $\Sigma_9^\text{H}$ is a complicated $9\times 9$ matrix.
Similarly to the odd case, the determinant of its leading order,
\begin{equation}
\begin{aligned}
\text{det}\Sigma_9^\text{H} = \frac{\ell(\ell+1)(\ell^2+\ell+4)}{256(2\di\,\tilde{\omega}-1)}
\Bigg\lbrace
& \big(\ell^2(\ell+1)^2-8 \big)\big(\tilde{\omega}^2-\tmT\big)
+ \frac{9(\ell-1)\ell(\ell+1)(\ell+2)}{2}\,\tG\\
& \
+ \bigg[64- \ell(\ell+1)\bigg(22 + \frac{15\ell(\ell+1)}{2}\bigg)\bigg]\,\di \,\tilde{\omega}\\
& \
- \ell^2(\ell+1)^2(\ell^2+\ell+5)
+2\ell(\ell+1) + 32\Bigg\rbrace\,,
\end{aligned}
\end{equation}
is non vanishing.
Therefore, it is diagonalizable, but we did not find a closed form expression for this purpose.
But, going to large $\ell$, one can extract the nine eigenvalues 
\begin{subequations}
\begin{align}
&
\tilde{\sigma}_2 = \di\,\tilde{\omega} -\frac{1}{2} - \frac{2(\ell^2+\ell-3)+5\di\,\tilde{\omega}-2(3\tilde{\omega}^2-\tmT)}{2(1-2\di\,\tilde{\omega})} + \mathcal{O}\left(\frac{1}{\ell}\right)\,,\\
&
\tilde{\sigma}_{3,4} = \frac{\ell}{2}- \frac{5}{4}\pm\sqrt{\frac{5}{2}}+ \mathcal{O}\left(\frac{1}{\ell}\right)\,,\\
&
\tilde{\sigma}_{5,6} = -\frac{\ell}{2}- \frac{7}{4}\pm\sqrt{\frac{5}{2}}+ \mathcal{O}\left(\frac{1}{\ell}\right)\,,\\
&
\tilde{\sigma}_{7,8} = \frac{\ell}{2} \pm \frac{\sqrt{2\ell}}{4} + \frac{15}{8}\pm\frac{9\tG -20\di\,\tilde{\omega}-243}{16\,\sqrt{2\ell}}+ \mathcal{O}\left(\frac{1}{\ell}\right)\,,\\
&
\tilde{\sigma}_{9,10} = -\frac{\ell}{2} \pm \di\,\frac{\sqrt{2\ell}}{4} + \frac{11}{8}\mp\frac{9\tG -20\di\,\tilde{\omega}-247}{16\,\sqrt{2\ell}}+ \mathcal{O}\left(\frac{1}{\ell}\right)\,,
\end{align}
\end{subequations}
The first one, $\tilde{\sigma}_2$ exactly matches the odd case~\eqref{eq_asym_odd_hor_z1} at this precision (note that the parameter  $\tG$ appears at the $\calO(\ell^{-1})$ order in $\tilde{\sigma}_{1..5}$).
So, and as advertised, we recover the same phenomenology for the propagating modes than in the odd sector, and the outgoing mode is dominant over the regular ingoing mode.

\section{Quasi-normal modes}
\label{sec_QNM}

Equipped with the asymptotic behavior of the perturbations, derived in the previous section, one can now implement the second step of the formalism developed in~\cite{Langlois:2021xzq,Roussille:2022vfa}, \emph{i.e.} use them as consistent boundary conditions to numerically derive the QNMs.
There exists a large class of numerical frameworks used to compute QNMs, and we will here adapt the continued-fraction method~\cite{Leaver:1985ax,Leaver:1990zz,Berti:2009kk,Pani:2013pma} to our case. This method was already applied to the specific case of a first-order matrix differential system in~\cite{Roussille:2023sdr}.
The main advantage of this method over, \emph{e.g.}, spectral methods or shooting methods, is that it is less sensitive to degenerate boundary conditions, such as the one we encounter at horizon.

\subsection{Continued-fraction method}
\label{subsec_QNM_gen}

A QNM is an eigenfrequency for the system~\eqref{eq_asym_1storder_odd} with physical boundary conditions, namely ingoing at horizon and outgoing at infinity.
Therefore the aim is to find the values of $\tilde{\omega}$ that render the system solvable, and to verify that such values correspond to the correct boundary conditions.

To do so, the continued-fraction method takes advantage from the simple fact that the  system~\eqref{eq_asym_1storder_odd}, and thus the sought solution, is smooth between infinity and the horizon.
Compactifying the spatial coordinate by using\footnote{Note here the first difference with the usual continued-fraction method (that uses the compact coordinate $\tilde{u} = 1-1/\rho$), due to the square-root dependence of the background metric~\eqref{eq_MTMG_ds2PG}.}
\begin{equation}
u \equiv \frac{\sqrt{\rho}-1}{\sqrt{\rho}}\,,
\end{equation}
this amounts in using the ansatz
\begin{equation}\label{eq_QNM_Xanz}
X_\text{odd}^\text{ansatz} = X_\text{odd}^\text{sing}(u) \sum_{n=0}^\infty f_n \,u^n\,, 
\end{equation}
where $X_\text{odd}^\text{sing}$ encompasses the singular behavior of $X_\text{odd}$ at the boundaries, and $\{f_n\}_{n\geq 0}$ are six-components arrays.
Indeed, having singled out the singular behavior of the solution, the remaining should be regular everywhere in $u\in ]0,1[$, thus the use of a series in $u$ as an ansatz.
Naturally, there is a direct correspondence between the fact that $X_\text{odd}^\text{ansatz}$ solves the system~\eqref{eq_asym_1storder_odd} and the convergence of the series.
Finding QNMs amounts to find the values of $\tilde{\omega}$ such that: (i)~$\sum_{n=0}^\infty f_n \,u^n$ has a convergence radius at least equal to 1; (ii)~$f_0$ and $\sum_{n=0}^\infty f_n$ are compatible with respectively ingoing boundary conditions at horizon, and outgoing ones at infinity.

In the case of MTMG, the singular sector of $X_\text{odd}$ is only given by its behavior at infinity, as the ingoing boundary condition at horizon amounts to $X_\text{odd} \sim $ constant, see Eq.~\eqref{eq_asym_odd_hor_zn}.
Therefore we simply set $X_\text{odd}^\text{sing}$ to be the singular behavior of  $X_\text{odd}$ at infinity, explicitly given by Eq.~\eqref{eq_app_Xodd_sing_infty} with the normalization constant $\chi_\infty$ set to unity.
Injecting such ansatz in the system~\eqref{eq_asym_1storder_odd}, it reduces to a (matrix) ten terms recurrence relation, that can be reduced to a three terms recurrence relation, using canonical Gaussian elimination,
\begin{subequations}\label{eq_QNM_recc_rel}
\begin{align}
&
\alpha_0\cdot f_1 + \beta_0\cdot f_0 = 0\,,\\
&
\alpha_n\cdot f_{n+1} + \beta_n\cdot f_n + \gamma_n\cdot f_{n-1} = 0\,,
\qquad
n \geq 1\,,
\end{align}
\end{subequations}
where $\{\alpha_n,\beta_n,\gamma_n\}_{n\geq 0}$ are $6\times6$ matrices.
The series $\sum_{n=0}^\infty f_n \,u^n$ obeying a recurrence relation of the type~\eqref{eq_QNM_recc_rel} has a convergence radius at least equal to 1 if and only if
\begin{equation}\label{eq_QNM_recc_sol}
M_0\cdot f_0 = 0\,,
\end{equation}
where the matrix $M_0$ is defined \emph{via} a continued-fraction algorithm
\begin{equation}\label{eq_QNM_recc_M0_def}
M_0 = \big(\alpha_0 \cdot R_0 + \beta_0\big)\,,
\qquad\text{with}\qquad
R_{n-1} = - \big(\alpha_n \cdot R_n + \beta_n\big)^{-1}\cdot\gamma_n\,,
\end{equation}
hence the name of the method.

For fixed value of $\{\ell,\tmT\}$, a given $\tilde{\omega}$ thus represents a quasi-normal mode if both the associated matrix $M_0$ has a vanishing determinant, and the corresponding vector of coefficients $f_0$ is coherent with the ingoing boundary condition at the horizon.
Note that although the fundamental condition~\eqref{eq_QNM_recc_sol} should in principle yield all QNMs, it is a known fact that the $n^\text{th}$ overtone is the most stable solution of the $n^\text{th}$ inversion of the continued-fraction relation~\eqref{eq_QNM_recc_M0_def}~\cite{Leaver:1985ax}.
This fact allowed for instance to compute extremely high overtones in GR by slightly adapting the algorithm~\cite{Nollert:1993zz,Berti:2004um,Konoplya:2004uk}.
Nevertheless, we will only focus on the fundamental mode, and thus the fundamental condition~\eqref{eq_QNM_recc_sol} will be largely sufficient for the scope of this work.
We let a deeper study of overtones in MTMG to further investigations.

\subsection{Results}
\label{subsec_QNM_results}

\begin{figure}[!h]
	\includegraphics{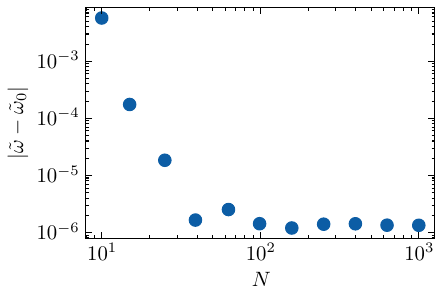}
	\caption{Difference between the computed value for the fundamental mode for several values of $N$ and its value computed at $N = 1100$. The observed convergence is proof that the mode obtained is indeed a QNM and not an artifact of the numerical method. One observes saturation at $10^{-6}$ precision, which corresponds to the threshold for convergence used in the numerical method.}
	\label{fig:convergence-QNM}
\end{figure}

Naturally, it is not possible to determine analytically the matrix $M_0$~\eqref{eq_QNM_recc_M0_def}.
Nevertheless, once $\{\ell,\tilde{\omega},\tmT\}$ are fixed, one can proceed numerically to a given precision, say $N$, and initiate the continued-fraction algorithm by taking arbitrarily $R_{N+1} = 0$,\footnote{There exist more subtle methods for the initiation of the algorithm, taking a motivated ansatz for $R_{N+1}$, see \emph{e.g.}~\cite{Nollert:1993zz}. Nevertheless, such subtleties are largely beyond the sought precision of the present work.} which gives a numerical estimate of $M_0$.
One can then find the values $\tilde{\omega}$ that put the determinant of such approximate $M_0$ to zero up to numerical precision, and thus could correspond to QNMs. Imposing that those values be stable under a change of the precision $N$ allows us to discriminate between ``true'' QNMs and spurious artifacts of the numerical method: see Fig.~\ref{fig:convergence-QNM} for a proof of convergence. Such a method was initially applied in~\cite{Roussille:2023sdr}.

Since we do not have explicit expressions for the behavior of perturbations at the horizon, we cannot extract an analytical form for the ingoing boundary condition. Therefore, we only require that the determinant of $M_0$ be zero and check that the corresponding kernel is of dimension 1.

\begin{figure}[!h]
\includegraphics{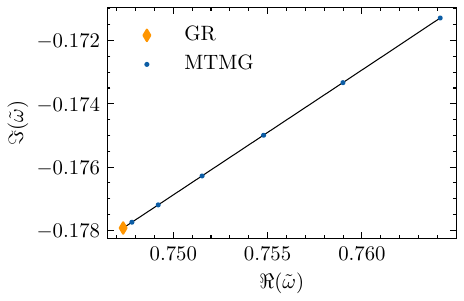}
\caption{Evolution of the real and imaginary parts of the odd parity fundamental QNM of MTMG, for $\tilde{\mu}_\mathrm{T}$ evolving between 0 and $0.3$ (blue dots correspond to increments of 0.05). In the $\tmT \to 0$ limit, the GR values are smoothly recovered.}
\label{fig_QNM}
\end{figure}

Focusing on the $\ell = 2$ case, we have implemented this practical method to track the fundamental QNM $\tilde{\omega}_0$, for different values of $\tmT$.
First, no QNM with positive imaginary part were found, at least in the range $\vert \Re(\tilde{\omega}\vert \leq 1$ and $\Im(\tilde{\omega}) \leq 1$.
This important results indicates that (to the extend of the domain we searched) black holes are stable in MTMG.
Then, when $\tmT = 0$, we found that the fundamental QNM is identical to the GR value $\tilde{\omega}_0^\mathrm{GR}$.
For a non-vanishing graviton mass, those QNMs differ from the RG ones, the limit $\tmT \to 0$ being regular, as clear from Fig.~\ref{fig_QNM}. For the range of values of $\tmT$ probed in this work, the fundamental mode $\tilde{\omega}_0$ is approximated up to numerical precision as
\begin{equation}
	\tilde{\omega}_0 \simeq \tilde{\omega}_0^\mathrm{GR} + (0.186592 + 0.073652\;\mathrm{i}) \times \tmT \,.
\end{equation}
This result is a new confirmation that, at the level of the theory, GR is a smooth limit of MTMG.
Moreover, it opens a new path of setting limit on the graviton mass, through a precise measure of the ringdown frequencies.

For the range of values of $\tmT$ considered, we also found that the fourth component of $X_\mathrm{odd}$, corresponding to $\ell^-$, was zero in all of the BH exterior up to numerical precision. To go further, we confirmed that imposing from the beginning $\ell^- = 0$ turned Eq.~\eqref{eq_asym_1storder_odd} into a new four-dimensional first-order ODE whose fundamental QNM has the same value $\tilde{\omega}_0$. Therefore, the perturbation of the Lagrange multiplier $\lambda^a$ can be consistently put to zero in the odd-parity sector of the perturbations.

\section{Conclusion and discussion}
\label{sec_concl}

This work investigates the phenomenology of gravitational ringdown within the Minimal Theory of Massive Gravity.
This theory enjoys the feature of massive gravitational waves, without containing additional polarizations, suffering from instability nor requiring screening mechanisms~\cite{DeFelice:2015hla,DeFelice:2015moy}.
Therefore, it offers a very convenient framework to test deviations from GR during the generation of gravitational radiation.
Starting from the black hole spacetimes derived in \cite{DeFelice:2018vza}, that mimic the usual black holes of GR, we applied the procedure~\cite{Langlois:2021xzq,Roussille:2022vfa,Roussille:2023sdr}, consisting in reducing the perturbation problem to a set of coupled first-order differential equations.
An asymptotic study of this system, at both horizon and spatial infinity, allows to construct consistent boundary conditions for the numerical derivation of QNMs.
Using the continued fraction method~\cite{Leaver:1985ax,Leaver:1990zz} and its adaptation to the computation of QNMs for a first-order system~\cite{Roussille:2023sdr}, we were able to find the fundamental QNM for different values of the graviton mass. We emphasize that such a result was made possible by the implementation of~\cite{Roussille:2023sdr}, since, up to our knowledge, there is no Schrödinger-like equation reformulation for BH perturbations in MTMG.

The main results of this study are that black holes in MTMG are stable, and that their (first) odd QNMs smoothly reduce to the GR values in the massless limit, while being different from GR ones when the graviton's mass is non-vanishing.
Therefore, we have confirmed that GR is a smooth limit of MTMG.
Moreover, those results open a new exciting possibility of  constraining MTMG with next-generation gravitational-wave detectors, such as LISA or ET\footnote{Note that establishing such a constraint would require the precise knowledge of the spin of the black hole, as its effect in GR can be degenerate with the effect of the graviton's mass~\cite{Leaver:1985}. However, there is no possible degeneracy with a putative charge of the black hole~\cite{Leaver:1990}, nor with the effect of a cosmological constant in GR~\cite{Zhidenko:2003wq,Berti:2009kk}.}.

The computation of QNMs was achieved only in the odd parity sector, and we let investigations of the (more cumbersome) even parity sector, together with the fundamental question of isospectrality, opened for future study.
Even more excitingly, now that we know that the black holes of MTMG are stable, we can aim at tackling the derivation of the full waveform.
This would notably require the subtle implementation of weak-field techniques, used to deal with the inspiral phase.
It would be particularly interesting to see how the mass term and the non-physical auxiliary fields enter the picture, and what would be the equivalent, in such Lorentz-violating theory, of the usual Bondi-like coordinate system, used to define observed quantities at spatial null infinity

\acknowledgments

It is a pleasure to thank E. Babichev, A. De Felice, \'E. Gourgoulhon, D. Langlois, S.~Mukohyama and K. Noui for enlightening discussions.
Both authors would like to express their gratitude to the  Institute for Basic Science (Daejeon, South Korea) for hosting them during a week of rich and fruitful discussions at the ``IBS CTPU-CGA Workshop on Modified Gravity'', at a late stage of this work.

\appendix

\section{The case of GR in Painlev\'e--Gullstrand coordinates}
\label{app_PG}

For the sake of clarity, we have not displayed lengthy expressions, notably during the study of the asymptotic regime performed in Sec.~\ref{sec_asymptotic}.
Nevertheless, and in order to provide a comprehensive overview of the method that was used, this Appendix applies the employed framework to the drastically simpler case of GR.
In order to be as close as possible to the rest of this work, we start from a Schwarzschild-de Sitter spacetime written in generalized Painlevé-Gullstrand (PG) coordinates, Eq.~\eqref{eq_MTMG_ds2PG}, and we only focus on odd parity modes.

For the sake of comparison, this section is organized as a mirror of the main study.
Sec.~\ref{app_PG_1st_order} presents the reduction of the problem to a first-order system, and corresponds to Sec.~\ref{subsec_asymptotic_reduction}.
Then, Sec.~\ref{app_PG_asymptotic} outlines the asymptotic study of the perturbations, similarly to the study performed in Sec.~\ref{subsec_asymptotic_odd}.

\subsection{Reduction to a first-order system}
\label{app_PG_1st_order}

Let us perturb the metric~\eqref{eq_MTMG_ds2PG} by selecting only the odd parity modes of Eq.~\eqref{eq_asym_pertmetric}
\begin{subequations}\label{eq_app_pert_def}
\begin{align}
& N = 1\,,
& &
\gamma_{rr} = 1\,,
\\
& N^r =\beta(r)\,,
& &
\gamma_{ra} = e^{-\di \omega t}\,h^-_{\ell,m}(r)\,\tilde{\varepsilon}_{ab}\,\partial^bY_L^M \,,\\
& N^a  =e^{-\di \omega t}\,n^-_{\ell,m}(r)\,\tilde{\varepsilon}^{ab}\,\partial_bY_L^M \,,
& &
\gamma_{ab} = r^2\omega_{ab} - e^{-\di \omega t}\,k^-_{\ell,m}(r)\,\tilde{\varepsilon}_{c(a}\nabla^c\partial_{b)}Y_L^M \,,
\end{align}
\end{subequations}
where we have shortened $\beta(r) = \sqrt{2M/r - \Lambda r^2/3}$, and we recall the notations introduced in Sec.~\ref{subsec_asymptotic_reduction}: $Y_L^M$ are the spherical harmonics, the 2-sphere metric is $\dd \Omega^2 = \omega_{ab}\dd \zeta^a\dd\zeta^b$ with compatible covariant derivative given by $\nabla_a$, and $\tilde{\varepsilon}_{ab} = \sqrt\omega \,\varepsilon_{ab}$ is the weighted anti-symmetric symbol.
As the background is static, we decompose on a time-Fourier basis and thus the three perturbations $\{n^-_{\ell,m},h^-_{\ell,m},k^-_{\ell,m}\}$ only depend on the radial coordinate $r$.
In the following, we concentrate on propagating modes $\ell \geq 2$, we use adimensioned variables $\rho = r/(2M)$ and $\tilde{\omega} = 2M\omega$, and we do not write the $(\ell,m)$ dependencies of the perturbation functions, as different $(\ell,m)$ modes decouple at linear level.

The three equations of motions $\mathcal{E}_\star \equiv \delta \mathcal{S}_\text{EH}/\delta \star$ are not independent but they are linked by the identity
\begin{equation}
\di\, \tilde{\omega} \rho \,\mathcal{E}_{n^-} + (\ell-1)(\ell+2)\,\mathcal{E}_{k^-}+\frac{1}{\rho^2}\frac{\dd}{\dd \rho}\bigg[\rho^3 \beta\,\mathcal{E}_{n^-}-\rho^2 \mathcal{E}_{h^-}\bigg] =0\,.
\end{equation}
This relation is nothing but the consequence of the gauge freedom of GR, and thus we can place ourselves in the usual Regge-Wheeler gauge~\cite{Regge:1957td} by setting $k^- = 0$.
In such gauge, and using the appropriate change of variables, the equations directly reduce to a system of two first-order differential equations
\begin{equation}\label{eq_app_1storder}
\frac{\dd X}{\dd \rho} + B X = 0\,,
\qquad
\text{with}\quad
X = 
\frac{1}{\rho}\begin{pmatrix}
\beta\,n^-+h^- \\
\di \,\tilde{\omega}\big[n^-+\beta\,h^-\big] + \frac{1-\beta^2}{r}\big[\beta\,n^-+h^-\big]
\end{pmatrix}\,,
\end{equation}
where the $B$-matrix reads
\begin{equation}\label{eq_app_Bexact}
B = \frac{1}{1-\beta^2}
\begin{pmatrix}
-\di\,\tilde{\omega}\,\beta & 1\\
V(\rho)-\tilde{\omega}^2 & -\di\,\tilde{\omega}\,\beta
\end{pmatrix}\,,
\end{equation}
with the canonical potential~\cite{Regge:1957td}
\begin{equation}
V(\rho) =\frac{\ell(\ell+1)\rho -3}{\rho^3}\big(1-\beta^2\big)\,.
\end{equation}
In order to recover the canonical second-order differential equation from this system, it is sufficient to perform the change of variables
\begin{equation}
r_\star = 2M \int_1^\rho\!\! \frac{\dd u}{1-\beta^2(u)}\,,
\qquad
\begin{pmatrix}
y_1 \\
y_2
\end{pmatrix}
= e^{-\di\,\omega \int^{r_\star}\!\dd u\, \beta(u)} \, X\,,
\end{equation}
which yields the usual Schrödinger-like equation
\begin{equation}\label{eq_app_Schro}
\frac{\dd^2 y_1}{\dd r_\star} + \bigg[\omega^2-\frac{V(r_\star)}{4M^2}\bigg]y_1 = 0\,.
\end{equation}
Naturally, when $\Lambda = 0$ $r_\star = r +2M \ln(r/2M-1)$ is the canonical Eddington–Finkelstein radial coordinate.

\subsection{Asymptotic regimes}
\label{app_PG_asymptotic}

Instead of studying directly the second-order differential equation~\eqref{eq_app_Schro}, let us apply the formalism developed in~\cite{Langlois:2021xzq,Roussille:2022vfa} and focus on the asymptotic structure of the first-order system~\eqref{eq_app_1storder}.
Naturally, this method is tailored for more intricate cases than GR, and applying it to GR is an unnecessary complication, but we recall that it is done here for illustrative purposes.
Hereafter, we work in an asymptotically flat spacetime by setting $\Lambda = 0$, and thus $\beta = \rho^{-1/2}$.

\subsubsection{Behavior at infinity}

A simple Taylor expansion of the $B$-matrix~\eqref{eq_app_Bexact} near spatial infinity yields
\begin{equation}
B = \begin{pmatrix}
0 & 1\\
- \tilde{\omega}^2 & 0
\end{pmatrix}
+ \frac{1}{\sqrt{\rho}}
\begin{pmatrix}
- \di\,\tilde{\omega} & 0\\
0 & - \di\,\tilde{\omega} 
\end{pmatrix}
+ \frac{1}{\rho}\begin{pmatrix}
0 & 1\\
- \tilde{\omega}^2 & 0
\end{pmatrix}
+ \calO\left(\frac{1}{\rho^{3/2}}\right)\,.
\end{equation}
The next step is to perturbatively diagonalize this expansion in order to solve it up to a given oder in $\rho$.
In this precise case, the diagonalization is trivially realized by the transition matrix
\begin{equation}
P_\infty = \begin{pmatrix}
\frac{\di}{\tilde\omega} & 1\\
1 & \di\, \tilde{\omega}
\end{pmatrix}\,.
\end{equation}
In terms of the new variables $Y_\infty = P_\infty^{-1}X$, it thus comes
\begin{equation}
\frac{\dd Y_\infty}{\dd \rho} = \di \,\tilde{\omega} \begin{pmatrix}
1 + \frac{1}{\sqrt{\rho}} + \frac{1}{\rho} + \calO\left(\frac{1}{\rho^{3/2}}\right) & \calO\left(\frac{1}{\rho^{3/2}}\right)\\
\calO\left(\frac{1}{\rho^{3/2}}\right) &  -1 - \frac{1}{\sqrt{\rho}} - \frac{1}{\rho} + \calO\left(\frac{1}{\rho^{3/2}}\right) 
\end{pmatrix} \,Y_\infty\,,
\end{equation}
which is easily solved.
In terms of the original variables $X$, it comes
\begin{equation}\label{eq_app_Xinf}
X \sim 
\begin{pmatrix}
1\\
-\di\,\tilde\omega
\end{pmatrix}
\, c_\infty^+\,e^{\di\,\tilde{\omega} \rho + 2\di\,\tilde{\omega}\sqrt{\rho}}\rho^{\di\,\tilde{\omega}}
+ \begin{pmatrix}
1\\
\di\,\tilde\omega
\end{pmatrix}
\, c_\infty^-\,e^{-\di\,\tilde{\omega} \rho + 2\di\,\tilde{\omega}\sqrt{\rho}}\rho^{-\di\,\tilde{\omega}}\,,
\end{equation}
where $\{c_\infty^\pm\}$ are two integration constants. 
It is easy to identify $c_\infty^+$ with the outgoing mode, propagating towards infinity, and $c_\infty^-$ with the ingoing mode, propagating from infinity.
Note the particular feature of the PG coordinate system, namely the unusual subdominant $e^{2\di\,\tilde{\omega}\sqrt{\rho}}$ behavior. 
The metric perturbations~\eqref{eq_app_pert_def} behave asymptotically as
\begin{equation}
\begin{aligned}
&
n^- \sim - c_\infty^+\,r^{1+2\di\,M\omega}\,e^{\di\,\omega r +  2\di\,\omega\sqrt{2M r}}
+  c_\infty^-\,r^{1-2\di\,M\omega}\,e^{-\di\,\omega r + 2 \di\,\omega\sqrt{2M r}}\,,\\
&
h^- \sim c_\infty^+\,r^{1+2\di\,M\omega}\,e^{\di\,\omega r +  2\di\,\omega\sqrt{2M r}}
+  c_\infty^-\,r^{1-2\di\,M\omega}\,e^{-\di\,\omega r + 2 \di\,\omega\sqrt{2M r}}\,,
\end{aligned}
\end{equation}
and thus are perfectly stable (at least in this coordinate system).

\subsubsection{Behavior at horizon} 

At horizon, let us work with $\theta \equiv (\rho -1)^{-1}$.
A Taylor expansion in $\theta \to \infty$ yields
\begin{equation}
\frac{\dd X}{\dd \theta} 
=
\begin{pmatrix}
- \di\,\tilde{\omega} & 1\\
- \tilde{\omega}^2 & - \di\,\tilde{\omega} 
\end{pmatrix}
\frac{X}{\theta}
+ \begin{pmatrix}
-\frac{\di\,\tilde{\omega}}{2} & 1\\
\ell^2+\ell-3- \tilde{\omega}^2 & -\frac{\di\,\tilde{\omega}}{2} 
\end{pmatrix}
\frac{X}{\theta^2}
+ \calO\left(\frac{1}{\theta^3}\right)\,.
\end{equation}
The transition matrix that diagonalize this system is constructed order by order and read at next-to-leading order
\begin{equation}
P_\text{H} = \begin{pmatrix}
\frac{\di}{\tilde\omega} & -\frac{\di}{\tilde\omega} \\
1 & 1
\end{pmatrix}
\cdot \Bigg[\mathbb{I}_2 
+ \frac{1}{\theta} \begin{pmatrix}
 0 & \frac{\di (\ell^2+\ell-3)}{3\tilde{\omega}(1-2\di\,\tilde{\omega})}\\
-\frac{\di (\ell^2+\ell-3)}{3\tilde{\omega}(1+2\di\,\tilde{\omega})} & 0
\end{pmatrix}
+ \calO\left(\frac{1}{\theta^2}\right)
\Bigg] \,.
\end{equation}
Recalling the transformation law~\eqref{eq_asym_odd_transfo_rule}, it comes in terms of the new variables $Y_\text{H} = P_\text{H}^{-1}X$
\begin{equation}
\frac{\dd Y_\text{H}}{\dd \theta} =  \begin{pmatrix}
-\frac{2\di\,\tilde{\omega}}{\theta} + \frac{\di (\ell^2+\ell-3-3\tilde{\omega}^2)}{2\tilde{\omega}\,\theta^2} + \calO\left(\frac{1}{\theta^3}\right) & \calO\left(\frac{1}{\theta^3}\right)\\
\calO\left(\frac{1}{\theta^3}\right) & -\frac{\di (\ell^2+\ell-3-\tilde{\omega}^2)}{2\tilde{\omega}\,\theta^2}  +\calO\left(\frac{1}{\theta^3}\right)
\end{pmatrix} \,Y_\text{H}\,,
\end{equation}
which is easily integrated.
The original variables $X$ thus reads
\begin{equation}\label{eq_app_Xhor}
X \sim 
\begin{pmatrix}
1\\
-\di\,\tilde\omega
\end{pmatrix}
\, c_\text{H}^+\,\big(\rho-1\big)^{2\di\,\tilde{\omega}}e^{\frac{\di (\ell^2+\ell-3-3\tilde{\omega}^2)}{2\tilde{\omega}}(\rho-1)}
+ \begin{pmatrix}
1\\
\di\,\tilde\omega
\end{pmatrix}
\, c_\text{H}^-\,e^{\frac{\di (\ell^2+\ell-3-\tilde{\omega}^2)}{2\tilde{\omega}}(\rho-1)}\,,
\end{equation}
where $\{c_\text{H}^\pm\}$ are two integration constants.
Using an Eddington–Finkelstein radial coordinate $\rho_\star = \rho +\ln(\rho-1)$, one can identify $c_\text{H}^+$ with the outgoing mode, propagating towards infinity, and $c_\text{H}^-$ with the ingoing one.
It is interesting to note that, contrary to the regime at infinity~\eqref{eq_app_Xinf}, the two modes do not behave similarly: the ingoing mode is smooth across the horizon when the outgoing mode has a mild singularity and a non-usual dispersion relation $\vert\vec{k}\vert = 2\omega$.
As discussed in Sec.~\ref{subsubsec_asymptotic_odd_hor}, those behaviors are just artifacts of the structure of the light-cone at horizon.
As for the original metric perturbations~\eqref{eq_app_pert_def}, they behave asymptotically as
\begin{equation}
\begin{aligned}
&
n^- \sim - \frac{\di\,c_\text{H}^+}{M\omega}\bigg(\frac{r}{2M}-1\bigg)^{4\di\,M\omega-1}
+ \frac{(5\di+4M\omega)M\omega-(\ell-1)(\ell+2)}{2M\omega(\di+4M\omega)}\,c_\text{H}^-\,,\\
&
h^- \sim \frac{\di\,c_\text{H}^+}{M\omega}\bigg(\frac{r}{2M}-1\bigg)^{4\di\,M\omega-1}
- \frac{(3\di-4M\omega)M\omega-(\ell-1)(\ell+2)}{2M\omega(\di+4M\omega)}\,c_\text{H}^-\,.
\end{aligned}
\end{equation}

\section{Lengthy expressions}\label{app_lengthy}
\allowdisplaybreaks

This Appendix collects the lengthy expressions that were not explicitely displayed in Sec.~\ref{sec_asymptotic}.

\subsection{Reduction of the odd sector}\label{app_lengthy_1storder_odd}

The quantities entering the first-order differential system~\eqref{eq_asym_1storder_odd} are given by
\begin{equation}\label{eq_app_Xodd_func_h}
X_\text{odd} =  \begin{pmatrix}
\frac{\dd}{\dd \rho}\bigg[n^- + \frac{h^-}{\rho^{1/2}}\bigg] \\
\frac{\dd k^-}{\dd \rho} - h^- \\
h^- \\
\frac{\rho^2\tmT}{\mathcal{X}-1}\,\ell^-\\
n^- + \frac{h^-}{\rho^{1/2}}\\
k^-
\end{pmatrix}\,,
\end{equation}
and
\begin{equation}\label{eq_app_Bodd}
B_\text{odd} = \begin{pmatrix}
0 & - \frac{(\ell-1)(\ell+2)}{2\rho^{5/2}}& \frac{(\ell-1)(\ell+2)}{2\rho^{5/2}} & 0 & - \frac{\ell(\ell+1)}{\rho^2} & \frac{(\ell-1)(\ell+2)}{\rho^{7/2}}\\
 - \frac{2\rho^{1/2}}{\rho-1} & \frac{3-2\rho-2\di\,\tilde{\omega}\rho^{3/2}}{\rho(\rho-1)} & - \frac{1}{\rho(\rho-1)} & \frac{2\rho}{\rho-1} & \frac{\rho^{-1/2}-2\di\,\tilde{\omega}\rho}{\rho-1} & b_1 \\
0 & 0 & \frac{2}{\rho} & 0 & 0 &  - \frac{(\ell-1)(\ell+2)}{2\rho^2}\\
-\di\,\tilde{\omega} &  \frac{(\ell-1)(\ell+2)(\rho-1)}{2\rho^3} &  \frac{2(\tilde{\omega}^2-\tmT)\rho^3-(\ell-1)(\ell+2)(\rho-1)}{2\rho^3} &  - \frac{2}{\rho} &  \frac{2\di\,\tilde{\omega}\rho^{3/2}-(\ell-1)(\ell+2)}{\rho^{5/2}} & \frac{(\ell-1)(\ell+2)(2-2\rho - \di\,\tilde{\omega}\rho^{3/2}}{2\rho^4}\\
-1 & 0 & 0 & 0 & 0 & 0 \\
0 & -1 & -1 & 0 & 0 & 0
\end{pmatrix}\,,
\end{equation}
with 
\begin{equation}
b_1 = \frac{(\tilde{\omega}^2-\tmT)\rho}{\rho-1} + \frac{5\di\,\tilde{\omega}}{2\rho^{1/2}(\rho-1)}- \frac{2}{\rho^2(\rho-1)}- \frac{(\ell-2)(\ell+3)}{2\rho^2}\,.
\end{equation}

\subsection{Reduction of the even sector}\label{app_lengthy_1storder_even}

The variables of the system~\eqref{eq_asym_1storder_even}, $X_\text{even} = (x_n)_{n=1..10}$, are related to the original even perturbations by
\begin{subequations}
\begin{align}
&
n^0 =
- \frac{x_1}{\rho^2} + 2 \,x_4- \frac{\ell(\ell+1)}{2\rho}\,x_5 + \frac{x_6}{\rho}
\,,\\
&
n^r =
\frac{2\rho+1}{4\rho^{5/2}}\,x_1 + \frac{\ell(\ell+1)}{4\rho^{1/2}}\,x_5 - \frac{x_6}{2\rho^{1/2}}
\,,\\
&
n^+ = 
x_3 - \frac{x_5}{\rho^{1/2}}
\,,\\
& 
h^0 =
- \frac{x_1}{2\rho^2}
\,,\\
&
h^+ =
x_5
\,,\\
&
k^0 =
x_1 + \frac{\ell(\ell+1)}{2}\,x_2
\,,\\
&
k^+ =
x_2
\,,\\
&
\nonumber
\ell^0 =
\frac{\mathcal{X}-1}{\tmT} \Bigg\lbrace
\frac{4(\tilde{\omega}^2-\tmT)\rho^3-3\tG \,\rho^3-4(\rho-1)-2\ell(\ell+1)}{8\rho^{7/2}}\,x_1\\
&\qquad\qquad\qquad\nonumber
- \frac{(\ell-1)\ell(\ell+1)(\ell+2)}{8\rho^{5/2}}\,x_2
+ \frac{\ell(\ell+1)(1+\di\,\tilde{\omega}\rho^{3/2}}{2\rho^2}\,x_3
+\frac{\ell(\ell+1)-2\di\,\tilde{\omega}\rho^{1/2}}{2\rho^{1/2}}\,x_4\\
&\qquad\qquad\qquad
+ \frac{2\rho-1-2\di\,\tilde{\omega}\rho^{3/2}}{4\rho^{5/2}}\,x_6
+ \frac{\ell(\ell+1)}{4\rho}\,x_8
- \rho^{1/2}\,x_9
\Bigg\rbrace
\,,\\
&
\nonumber
\ell^r =
\frac{\mathcal{X}-1}{\tmT} \Bigg\lbrace
\frac{(2\rho-1)\di\,\tilde{\omega}\rho^{3/2}-2(\rho-1)-\ell(\ell+1)\rho}{4\rho^4}\,x_1
- \frac{(\ell-1)\ell(\ell+1)(\ell+2)}{8\rho^3}\,x_2\\
&\qquad\qquad\qquad
+ \frac{\ell(\ell+1)}{4\rho^{5/2}}\,x_3
+ \frac{2\rho-1-2\di\,\tilde{\omega}\rho^{5/2}}{4\rho^3}\,x_6
+ \frac{\ell(\ell+1)}{4\rho^{1/2}}\,x_8
- x_9 \Bigg\rbrace
\,,\\
&
\ell^+ =
\frac{\mathcal{X}-1}{\tmT}\,x_{10}
\,.
\end{align}
\end{subequations}
As for the matrix $B_\text{even}$, it explicitly reads
\begin{equation}\label{eq_app_Beven}
B_\text{even} =
\begin{pmatrix}
- \frac{1}{2\rho} & 0 & 0 & 0 & \frac{\ell(\ell+1)}{2} & 0 & 0 & 0 & 0 & 0 \\
0 & 0 & 0 & 0 & -2 & 0 & -1 & 0 & 0 & 0 \\
\frac{1- 2\rho}{4\rho^{5/2}} & 0 & 0 & 0 & b_2 & \frac{1}{2\rho^{1/2}} & 0 & -1 & 0 & 0\\
\frac{2(\ell^2+\ell+2)\rho-3}{8\rho^4} & \frac{(\ell-1)\ell(\ell+1)(\ell+2)}{8\rho^3} & - \frac{\ell(\ell+1)}{4\rho^{5/2}} & 0 & b_3 & - \frac{1}{2\rho^2} & 0 & - \frac{\ell(\ell+1)}{4\rho^{1/2}} & 0 & 0\\
- \frac{1}{2\rho^2} & - \frac{(\ell-1)(\ell+2)}{2\rho^2} & 0 & 0 & \frac{2}{\rho} & 0 & 0 & 0 & 0 & 0 \\
\frac{3-2(\ell^2+\ell+2)\rho}{4\rho^3} & -\frac{(\ell-1)\ell(\ell+1)(\ell+2)}{4\rho^2} & \frac{\ell(\ell+1)}{2\rho^{3/2}} & 0 & b_4 & 0 & 0 & 0 & 0 & 0\\
- \frac{1}{2\rho^2} & b_5 & \frac{\rho^{-1/2}-2\di\,\tilde{\omega}\rho}{\rho-1} & - \frac{2\rho}{\rho-1} & - \frac{4}{\rho} & 0 & b_6 & - \frac{2\rho^{1/2}}{\rho-1} & 0 & \frac{2\rho^3}{\rho-1}\\
\frac{3-\di\,\tilde{\omega}\rho^{3/2}}{2\rho^{7/2}} & \frac{(\ell-1)(\ell+2)(2-\di\,\tilde{\omega}\rho^{3/2})}{2\rho^{7/2}} & 0 & -\rho^{-3/2} & b_7 & - \frac{1}{2\rho^{5/2}} & b_8 & 0 & 0 & 0\\
b_9 & \frac{(\ell-1)\ell(\ell+1)(\ell+2)(1-3\rho)}{16\rho^5} & b_{10} & - \frac{\ell(\ell+1)}{4\rho^2} & b_{11} & \frac{6\rho-5}{8\rho^4} & 0 & - \frac{\ell(\ell+1)}{4\rho^{5/2}} & \frac{1}{2\rho} & \frac{\ell(\ell+1)}{4}\\
b_{12} & b_{13} & b_{14} & \frac{2\rho-1}{\rho^4} & - \frac{\tmT}{\rho^2} & b_{15} & b_{16} & b_{17} & - \frac{1}{\rho^2} & 0
\end{pmatrix}\,,
\end{equation}
with
\begin{subequations}
\begin{align}
&
b_2 = \di\,\tilde{\omega} - \frac{\ell(\ell+1)}{4\rho^{1/2}}\,,\\
&
b_3 = \frac{\ell(\ell+1)(1+2\di\,\tilde{\omega}\rho^{5/2})}{8\rho^3}\,,\\
&
b_4 = \frac{\ell(\ell+1)(6\rho-1)}{4\rho^2}\,,\\
&
b_5 = \frac{(\tilde{\omega}^2-\tmT)\rho}{\rho-1} + \frac{5\di\,\tilde{\omega}}{2\rho^{1/2}(\rho-1)} + \frac{2(\rho-2)}{\rho^2(\rho-1)}\,,\\
&
b_6  = - \frac{-2\di\,\tilde{\omega}}{\rho^{1/2}(\rho-1)}+\frac{3-2\rho}{\rho(\rho-1)}\,,\\
&
b_7 = - \frac{(\ell-1)(\ell+2)}{2\rho^{5/2}}\,,\\
& 
b_8 = \frac{2\di\,\tilde{\omega}}{\rho}-\frac{\ell(\ell+1)}{4\rho^{5/2}}\,,\\
&
b_9 = \frac{9\,\tG}{16\rho^2} - \frac{3\di\,\tilde{\omega}}{4\rho^{7/2}}- \frac{(\ell^2+\ell+2)(3\rho-1)}{8\rho^5} + \frac{3(8\rho-1)}{16\rho^6}\,,\\
&
b_{10} = \frac{\ell(\ell+1)(6\rho-1-2\di\,\tilde{\omega}\rho^{5/2})}{8\rho^{9/2}}\,,\\
&
b_{11} = \frac{\ell(\ell+1)(1-4\rho+4\di\,\tilde{\omega}\rho^{5/2})}{16\rho^5} \,,\\
&
b_{12} = - \frac{2\rho+3}{4\rho^{9/2}}\,\di\,\tilde{\omega} + \frac{\ell(\ell+1)}{4\rho^5} - \frac{3(\rho-1)}{2\rho^6}\,,\\
&
b_{13} = \frac{(\ell-1)(\ell+2)}{8\rho^6}\bigg[\ell(\ell+1)\rho-8(\rho-1)-4\di\,\tilde{\omega}\rho^{3/2}\bigg]\,,\\
&
b_{14} = \frac{2\di\,\tilde{\omega}}{\rho^3}+ \frac{8-\ell(\ell+1)}{4\rho^{9/2}}\,,\\
&
b_{15} = \frac{\di\,\tilde{\omega}}{2\rho^{5/2}} + \frac{2\rho-3}{4\rho^5}\,,\\
&
b_{16} = \frac{(\ell-1)(\ell+2)(\rho-1)}{2\rho^5}\,,\\
&
b_{17} =- \frac{\di\,\tilde{\omega}}{\rho^2}  - \frac{\ell(\ell+1)}{4\rho^{5/2}}\,.
\end{align}
\end{subequations}

\subsection{Asymptotic behavior of the odd parity perturbations at infinity}

At spatial infinity, the odd parity perturbations behave at leading order as
\begin{subequations}\label{eq_app_asym_odd_infty}
\begin{align}
&
n^- \sim
- \frac{3(\ell-1)(\ell+2)\,\di\,\tilde{\omega}}{4(\tilde{\omega}^2-\tmT)}\,\frac{z_+^\text{prop}+z_-^\text{prop}}{\rho^{3/2}}
+ \frac{z_1}{\rho^\ell} + \frac{z_2}{4\rho^{\ell+3/2}} + \frac{z_3}{4}\,\rho^{\ell-1/2} + z_4\,\rho^{\ell+1}\,,\\
&
h^- \sim \nonumber
- \frac{(\ell-1)(\ell+2)\,\di}{2\sqrt{\tilde{\omega}^2-\tmT}}\,\frac{z_+^\text{prop}-z_-^\text{prop}}{\rho}
+ \frac{2(\ell-1)(\ell+2)(5\ell+4)}{3(4\ell+5)(\tilde{\omega}^2-\tmT)}\,\frac{z_1}{\rho^{\ell-5/2}}\\
&
\qquad\qquad
- \frac{z_2}{4\rho^{\ell+1}} - \frac{z_3}{4}\,\rho^{\ell} + \frac{2(\ell-1)(\ell+2)(5\ell+1)}{3(4\ell-1)(\tilde{\omega}^2-\tmT)}\,z_4\,\rho^{\ell-3/2}\,,\\
&
k^- \sim \nonumber
z_+^\text{prop}-z_-^\text{prop}
- \frac{(20\ell^2+38\ell+23)}{3(4\ell+5)(\tilde{\omega}^2-\tmT)}\,\frac{z_1}{\rho^{\ell-3/2}}\\
&
\qquad\qquad
+ \frac{z_2}{2(\ell+2)\rho^{\ell}} - \frac{z_3}{2(\ell-1)}\,\rho^{\ell+1} + \frac{(20\ell^2+2\ell+5)}{3(4\ell-1)(\tilde{\omega}^2-\tmT)}\,z_4\,\rho^{\ell-1/2}\,,\\
&
\ell^- \sim \frac{\mathcal{X}-1}{\tmT}\bigg[\di\,\tilde{\omega}\,\frac{z_1}{\rho^{\ell+2}} - \frac{\tilde{\omega}^2-\tmT}{4(\ell+2)}\,\frac{z_2}{\rho^{\ell+2}} +\frac{\tilde{\omega}^2-\tmT}{4(\ell-1)}\,z_3\,\rho^{\ell-1} +\di\,\tilde{\omega}\,z_4\,\rho^{\ell-1}\bigg]
\,.
\end{align}
\end{subequations}
where $\{z_\pm^\text{prop}\}$ are defined in~\eqref{eq_asym_odd_z_infty} and $\{z_n\}_{n=1..4}$ are four constants corresponding to the static modes.

Selecting only the outgoing modes $z_+^\text{prop}$, the intermediate variable $X_\text{odd}$~\eqref{eq_app_Xodd_func_h} thus behaves as
\begin{equation}\label{eq_app_Xodd_sing_infty}
X_\text{odd} \sim 
\begin{pmatrix}
\frac{(\ell-1)(\ell+2)}{2\rho^{3/2}}\\
\frac{1+\di\,\tilde{\omega}\rho^{1/2}}{2} + \frac{\di\,\sqrt{\tilde{\omega}-\tmT}}{4}\,(4\rho+1)\\
- \frac{\di}{2}\,\frac{(\ell-1)(\ell+2)}{\sqrt{\tilde{\omega}-\tmT}\,\rho}\\
0 \\
- \frac{\di}{2}\,\frac{(\ell-1)(\ell+2)}{\sqrt{\tilde{\omega}-\tmT}\,\rho^{3/2}}\\
\frac{4\rho-1}{4} + \frac{\di}{2}\frac{1+\di\,\tilde{\omega}\rho^{1/2}}{\sqrt{\tilde{\omega}-\tmT}}
\end{pmatrix} 
\,\chi_\infty
\,  e^{\di\,\sqrt{\tilde{\omega}^2-\tmT}\,\rho + 2\di\,\tilde{\omega}\sqrt{\rho}} \,\rho^{\frac{\di}{2} \frac{2\tilde{\omega}^2-\tmT}{\sqrt{\tilde{\omega}^2-\tmT}}}\,,
\end{equation}
where $\chi_\infty$ is a normalization constant.
Note that $x_4$, corresponding to $\ll^-$, does not propagate at infinity.

\bibliography{ListeRef_MTMG_pert.bib}

\end{document}